\documentclass{aa}  

\usepackage[dvipsnames]{xcolor}
\usepackage{graphicx}
\usepackage{multirow}
\usepackage[shortlabels]{enumitem}
\usepackage{ulem}
\usepackage{txfonts}
\usepackage{subfigure}
\usepackage{placeins}
\usepackage[
  breaklinks = true,
  colorlinks = true,
  urlcolor   = blue,
  citecolor  = blue,
  linkcolor  = blue,
]{hyperref}
\graphicspath{ {.} }

\newcommand{\Halpha}{{H$\alpha$}}

\begin{document} 

   \title{Thin coronal jets and plasmoid-mediated reconnection}
   \subtitle{Insights from Solar Orbiter observations and Bifrost simulations}
   \titlerunning{Thin coronal jets and plasmoid-mediated reconnection}

   \author{Daniel N\'obrega-Siverio\inst{1,2,3,4}
          \and
          Reetika Joshi\inst{3,4,5,6}
          \and
          Eva Sola-Viladesau\inst{2}
          \and
          David Berghmans\inst{7}
          \and
          Daye Lim\inst{8,7}
          }     
   \institute{Instituto de Astrof\'isica de Canarias, E-38205 La Laguna, Tenerife, Spain\\
    \email{dnobrega@iac.es}
    \and
    Universidad de La Laguna, Dept. Astrof\'isica, E-38206 La  Laguna, Tenerife, Spain
    \and
    Rosseland Centre for Solar Physics, University of Oslo, PO Box 1029 Blindern, 0315 Oslo, Norway
    \and
    Institute of Theoretical Astrophysics, University of Oslo, PO Box 1029 Blindern, 0315 Oslo, Norway
    \and
    Department of Physics and Astronomy, George Mason University, Fairfax, VA 22030, USA 
    \and
    Heliophysics Science Division, NASA Goddard Space Flight Center, Greenbelt, MD 20771, USA
    \and    
    Solar-Terrestrial Centre of Excellence–SIDC, Royal Observatory of Belgium, Ringlaan -3- Av. Circulaire, 1180 Brussels, Belgium
    \and
    Centre for mathematical Plasma Astrophysics, Department of Mathematics, KU Leuven, Celestijnenlaan 200B, 3001 Leuven, Belgium
    }

   \date{Received April 30, 2025; accepted August 18, 2025}

\abstract
{Coronal jets are ubiquitous, collimated million-degree ejections that contribute to the energy and mass supply of the upper solar atmosphere and the solar wind.
Solar Orbiter provides an unprecedented opportunity to observe fine-scale jets from a unique vantage point close to the Sun.}
{We aim to uncover thin jets originating from Coronal Bright Points (CBPs) and investigate observable features of plasmoid-mediated reconnection.}
{We analyze eleven datasets from the High Resolution Imager 174~\AA\ of the Extreme Ultraviolet Imager (HRI$_{\mathrm{EUV}}$) onboard Solar Orbiter, focusing on narrow jets from CBPs and signatures of magnetic reconnection within current sheets and outflow regions.
To aid interpretation, we compare the observations with radiative-MHD simulations of a CBP conducted with the Bifrost code.}
{We have identified thin coronal jets originating from CBPs with widths ranging from 253~km to 706~km: scales that could not be resolved with previous EUV imaging instruments.
Remarkably, these jets are 30–85\% brighter than their surroundings and can extend up to 22~Mm while maintaining their narrow form.
In one of the datasets, we directly identify plasmoid-mediated reconnection through the development within the current sheet of a small-scale plasmoid that reaches a size of 332~km and propagates at 40~km~s$^{-1}$.
In another dataset, we infer indirect traces of plasmoid-mediated reconnection through the intermittent boomerang-like pattern that appears in the outflow region.
The simulation self-consistently produces a current sheet and small-scale plasmoids similar to those observed, whose synthetic HRI$_{\mathrm{EUV}}$ emission reproduces both direct imprints within the current sheet and intermittent patterns in the outflow region associated with their ejection.}
{Our findings highlight Solar Orbiter's unique capability to capture narrow jets and sub-megameter-scale plasmoid-mediated reconnection signatures in the corona, motivating future statistical studies to assess the role of such fine-scale phenomena in coronal dynamics and solar wind formation.}

\maketitle

\section{Introduction}\label{s:introduction}
The Extreme Ultraviolet Imager \citep[EUI;][]{Rochus_etal:2020} onboard Solar Orbiter \citep[][]{Muller_etal:2020} is 
transforming our understanding of the solar atmosphere by revealing a wide variety of transient small-scale events as 
well as spatially and temporally resolving the fine structure of the corona.
Representative examples are EUV brightenings in the quiet Sun also known as campfires \citep[e.g.,][]{Berghmans_etal:2021,Zhukov_etal:2021,Panesar_etal:2021,Kahil_etal:2022,Huang_etal:2023,Dolliou_etal:2023,Dolliou_etal:2024,Nelson_etal:2023,
Nelson_etal:2024,Narang_etal:2025}; tiny inverted Y-shaped coronal jets with energies within the nanoflare and picoflare range \citep[e.g.,][]{Mandal_etal:2022,Chitta_etal:2023,Panesar_etal:2023,Shi_etal:2024,Chitta_etal:2025}; nanojets \citep{Gao_etal:2025}; EUV upflow-like events \citep{Duan_etal:2025};
bright dot-like features with sizes of $0.3-0.6$~Mm 
in emerging regions \citep{Tiwari_etal:2022} and active regions \citep{Mandal_etal:2023}; coronal oscillations \citep[e.g.,][]{Lim_etal:2024,
Shrivastav_etal:2024,Shrivastav_etal:2025,Meadowcroft_etal:2024,
Meadowcroft_Nakariakov:2025}; fast bidirectional propagating brightenings in arch filament systems 
\citep{Chen_etal:2024}; among others.

EUI's capabilities also make Solar Orbiter particularly promising for unraveling signatures of magnetic reconnection related to Coronal Bright Points (CBPs).
CBPs are small-scale, million-Kelvin loop structures that exhibit strong X-ray and/or EUV emission 
over durations of hours to days \citep[e.g.,][]{Madjarska:2019,Kraus_etal:2023}.
These phenomena are considered fundamental building blocks of the solar atmosphere for several reasons.
First, after active regions, CBPs are the primary contributors to high-energy radiation across the solar disk 
\citep{Mondal_etal:2023}.
Second, they are ubiquitous and nearly uniformly scattered across the entire solar disk, irrespective of the solar cycle 
phase \citep{Madjarska:2019}.
Third, CBPs serve as source regions for standard and blowout/breakout jets \citep[e.g.,][]{Hong_etal:2014,Sterling_etal:2015,Panesar_etal:2018,Kumar_etal:2018,Kumar_etal:2019a,Madjarska_etal:2022}.
Additionally, CBPs are often associated with fan-spine topologies, where magnetic reconnection occurs between closed 
and open field lines (a process usually referred to as interchange reconnection). 
This configuration is key to understanding the low-atmospheric origins of solar wind switchbacks \citep[e.g.,][]{Fargette_etal:2021,Bale_etal:2021,Wyper_etal:2022,Gannouni_etal:2023,Touresse_etal:2024}.

Recent studies have begun to capture the small-scale details of CBPs using HRI$_{\mathrm{EUV}}$
 observations.
For instance, \cite{Cheng_etal:2023} identify tiny bright 
blobs propagating along the outer spine and fan surface of the CBP with velocities from 30 to 210~km~s$^{-1}$ and 
lifetimes varying from 5 s (limited by the instrument cadence) to 105 s.
\cite{Petrova_etal:2024} report rotational motions traveling along the 30~Mm outer spine of a CBP, with speeds
between 136 and 160 km~s$^{-1}$.
Furthermore, HRI$_{\mathrm{EUV}}$ observations have also been used to better understand both standard and blowout jets from CBPs, 
as well as the transition between these two types \citep[e.g.,][]{Mandal_etal:2022,Long_etal:2023}.
Simultaneously, significant theoretical efforts have been made to accurately model CBPs and reproduce their EUV synthetic observables 
taking advantage of the state-of-the-art Bifrost code \citep{Gudiksen_etal:2011}.
For example, in the 2D model by \cite{Nobrega-Siverio_Moreno-Insertis:2022}, the authors focus on ejections associated 
with CBPs, which were characterized by two distinct stages: a main stage, featuring continuous reconnection with bursty 
behavior that produced a narrow coronal jet; and an eruptive stage, where a small-scale flux emergence episode completely 
disrupted the fan-spine topology, resulting in a broad, large coronal jet reaching temperatures up to 10 MK, accompanied 
by a chromospheric surge.
In following 2D studies, \cite{Faerder_etal:2024a,Faerder_etal:2024b} analyze the properties of plasmoids formed and ejected in the current sheet of the fan-spine topology. 
Their findings reveal plasmoids with scales of approximately $0.2-0.5$~Mm and lifetimes of $10-20$~s, suggesting that HRI$_{\mathrm{EUV}}$ has the potential to detect such small-scale, fast-moving plasmoids.
Moving to 3D models, \cite{Nobrega-Siverio_etal:2023} demonstrated the sustained heating of CBPs over several hours, 
finding that the main heating of CBPs occurs at their loop footpoints through
a braiding-like mechanism, with a secondary contribution from the heating at the null point. 
These models successfully reproduce many observational features of CBPs and their associated ejections.
However, a direct comparison with high-resolution, high-cadence HRI$_{\mathrm{EUV}}$ observations is still needed to better understand CBPs and related jets, as well as the signatures of fast magnetic reconnection involving plasmoids and shocks.

In this paper, we investigate several open questions related to the fine structure of CBPs and their associated jets. 
These include how narrow the jets can be, the presence of direct signatures of tearing instability in the current sheet associated with the CBP's null point, and indirect signatures of magnetic reconnection after plasmoids exit the current sheet and interact with the pre-existing magnetic field.
The layout of this paper is as follows. In Sect.~\ref{s:methods}, we give a description of the observations and numerical 
model used to compare the HRI$_{\mathrm{EUV}}$ data.
In Sect.~\ref{s:results_jets}, we present the properties of the observed CBP and related thin jets.
Then, in Sect.~\ref{s:results_reconnection}, we show the signatures of plasmoid-mediated magnetic reconnection and the comparison with the simulation. 
Finally, Sect.~\ref{s:discussion} contains the main conclusions and discussion.

\section{Methods}\label{s:methods}

\subsection{Observations}\label{s:methods_obs}
We use high-resolution, high-cadence EUV data from Solar Orbiter \citep{Muller_etal:2020}.
The data are obtained from the 174~\AA\ passband of the High Resolution Imager telescope that is part of the 
Extreme Ultraviolet Imager \citep[HRI$_{\mathrm{EUV}}$;][]{Rochus_etal:2020}.
For this study, we reviewed the publicly available catalog\footnote{\url{https://sidc.be/EUI/data/states/}} 
\citep[data release 6.0,][]{Kraaikamp_etal:2023} to select HRI$_{\mathrm{EUV}}$ level-2 observations
of CBPs from 2021 to 2023 that feature a jet spine.
We prioritized those taken at 
$\alpha_{\mathrm{Earth}}$, the longitudinal separation between the Earth and Solar Orbiter, between 15$^{\circ}$ and 55$^{\circ}$ to explore a range slightly offset from the Solar Dynamics Observatory
\citep[SDO;][]{Pesnell_etal:2012}, while maintaining sufficient overlap for comparative analysis.
An additional key criterion for the selection was the presence of a sufficiently long observational sequence; this way, we 
included only observations with at least 100 frames. 
In total, we have found eleven clear datasets whose details are summarized in Table~\ref{tab:solo_observations}.
Concerning the data processing, the images of each dataset are co-aligned to the first image of the series, taking the shift and rotation information of the metadata provided in the image headers and applying spline interpolation \citep[see also][]{Chitta_etal:2022,Mandal_etal:2022,Lim_etal:2024,Petrova_etal:2024}. 
We have also accounted for solar rotation in the co-alignment.
No further processing is applied to enhance the HRI$_{\mathrm{EUV}}$ images shown in this paper.

We have also employed 171 \AA\ data from the Atmospheric Imaging Assembly \citep[AIA;][]{Lemen_etal:2012} and 
photospheric magnetograms from 
the Helioseismic and Magnetic Imager \citep[HMI;][]{Scherrer_etal:2012} onboard SDO for comparison and context purposes.

\begin{table*}
    \caption{
    Details of the eleven Solar Orbiter HRI$_{\mathrm{EUV}}$ datasets analyzed. 
    }
    \label{tab:solo_observations}
    \centering
    \setlength\doublerulesep{0.5pt} 
    \begin{tabular}{ c c c c c c c c c c }
        \hline
        \hline
        \rule{0pt}{2.5ex}
        Case & Date & t$_{\mathrm{start}}$ (UT) & t$_{\mathrm{end}}$ (UT) & $\Delta$t (s) & D$_{\mathrm{Sun}}$ (au) & 
        Pixel size (km) & t$_{\mathrm{exp}}$ (s) & $\alpha_{\mathrm{Earth}}$ (deg) & (X,Y)\\
        \hline
        \rule{0pt}{2.5ex}
        01 & 2023-04-07 & 04:20:00 & 05:50:00 & 3  & 0.30 & 108 & 1.65 & 41.24 & (830", -298")\\ 
        \rule{0pt}{2.5ex}
        02 & 2023-04-05 & 04:00:05 & 04:19:15 & 10 & 0.32 & 115 & 3.65 & 29.79 & (471", 335")\\ 
           & 2023-04-05 & 04:19:35 & 04:45:55 & 10 & 0.32 & 115 & 3.65 & 29.79 & (478", 335")\\
        \rule{0pt}{2.5ex}
        03 & 2023-04-04 & 04:32:08 & 06:18:03 & 3  & 0.33 & 119 & 1.65 & 24.58 & (480", -130")\\ 
           & 2023-04-04 & 06:18:08 & 06:47:59 & 5  & 0.33 & 119 & 3.65 & 24.58 & (509", -130")\\ 
        \rule{0pt}{2.5ex}
        04 & 2022-03-22 & 10:39:40 & 16:29:40 & 30 & 0.33 & 119 & 2.80 & 48.14 & (680", 35")\\ 
        \rule{0pt}{2.5ex}
        05 & 2022-03-18 & 10:10:00 & 11:09:55 & 5  & 0.37 & 133 & 2.80 & 29.87 & (470", -505")\\ 
        \rule{0pt}{2.5ex}
        06 & 2022-03-17 & 03:18:00 & 03:28:36 & 3  & 0.38 & 137 & 1.65 & 25.98 & (545", 530")\\ 
           & 2022-03-17 & 03:28:42 & 04:02:57 & 3  & 0.38 & 137 & 1.65 & 25.98 & (551", 530")\\
        \rule{0pt}{2.5ex}
        07 & 2022-10-26 & 19:00:00 & 19:23:55 & 5  & 0.42 & 151 & 1.65 & 48.62 & (-198", 238")\\ 
        \rule{0pt}{2.5ex}
        08 & 2022-10-29 & 19:00:00 & 19:29:55 & 5  & 0.46 & 166 & 1.65 & 40.79 & (-250", 485")\\ 
           & 2022-10-29 & 22:00:00 & 22:29:55 & 5  & 0.46 & 166 & 1.65 & 40.79 & (-223", 485")\\ 
        \rule{0pt}{2.5ex} 
        09 & 2023-10-24 & 11:45:00 & 13:44:50 & 10 & 0.46 & 166 & 5.00 & 33.94 & (-590", 110")\\ 
        \rule{0pt}{2.5ex}
        10 & 2023-10-29 & 06:50:02 & 07:29:52 & 10 & 0.53 & 191 & 6.65 & 25.11 & (70", 550")\\ 
        \rule{0pt}{2.5ex}
        11 & 2021-09-14 & 07:03:22 & 07:46:02 & 20 & 0.59 & 212 & 0.70 & 47.76 & (120", 345")\\ 
        \hline
    \end{tabular}
    \tablefoot{The table contains columns for: the case number; observation date; start and end times, t$_{\mathrm{start}}$ and 
    t$_{\mathrm{end}}$, respectively; observation cadence, $\Delta$t; the distance of Solar Orbiter to the Sun, 
    D$_{\mathrm{Sun}}$; the corresponding pixel size; the exposure time, t$_{\mathrm{exp}}$; the longitudinal separation between the Earth and Solar Orbiter, 
    $\alpha_{\mathrm{Earth}}$; and the Solar coordinates (X,Y) of the event of interest as seen from Earth. 
    For cases 02, 03, 06, and 08, Solar Orbiter takes observations of the same target at different time windows.}
\end{table*}

\subsection{Numerical experiment}\label{s:methods_sim}
To provide theoretical support, we use a radiative-MHD numerical experiment of CBP performed 
with the Bifrost code \citep{Gudiksen_etal:2011}.
The experiment, presented and detailed in \cite{Nobrega-Siverio_Moreno-Insertis:2022}, is a 2D simulation based on a fan-spine magnetic topology with a null point at coronal heights (8~Mm above the solar surface), mimicking the configurations typically 
inferred from CBP observations  \citep[e.g.,][]{Zhang_etal:2012, Mou_etal:2016,Galsgaard_etal:2017,JoshiR_etal:2020, Madjarska_etal:2021,Cheng_etal:2023,Petrova_etal:2024}.
The experiment self-consistently demonstrates how the hot loops of a CBP can be formed solely through the action of stochastic granular motions, with magnetic reconnection in the corona mediating the process.
The reconnection exhibits both intermittent and oscillatory behavior, leading to the ejection of multiple plasmoids and sustained megakelvin jetting activity.
The simulation is performed at high spatial resolution ($\approx16$~km in both directions), allowing us to resolve the fine-scale structure of the coronal ejections and associated reconnection features.
The realism of this radiative-MHD experiment has proven valuable for interpreting chromospheric observations underneath fan-spine topologies \citep{Bose_etal:2023,Bhatnagar_etal:2025b}, and for producing synthetic EUV observables of plasmoids in support of upcoming missions such as the Multi-slit Solar Explorer \citep[MUSE;][]{Cheung_etal:2022,De-Pontieu_etal:2022}.
In this case, to compare with Solar Orbiter observations, we compute synthetic HRI$_{\mathrm{EUV}}$ emission images 
\cite[see][for details]{Nobrega-Siverio_Moreno-Insertis:2022,Nobrega-Siverio_etal:2023,Faerder_etal:2024b}.

\section{Thin coronal jets from CBPs}\label{s:results_jets}

\begin{figure*}[!ht]
   \centering
   \includegraphics[width=0.80\textwidth]{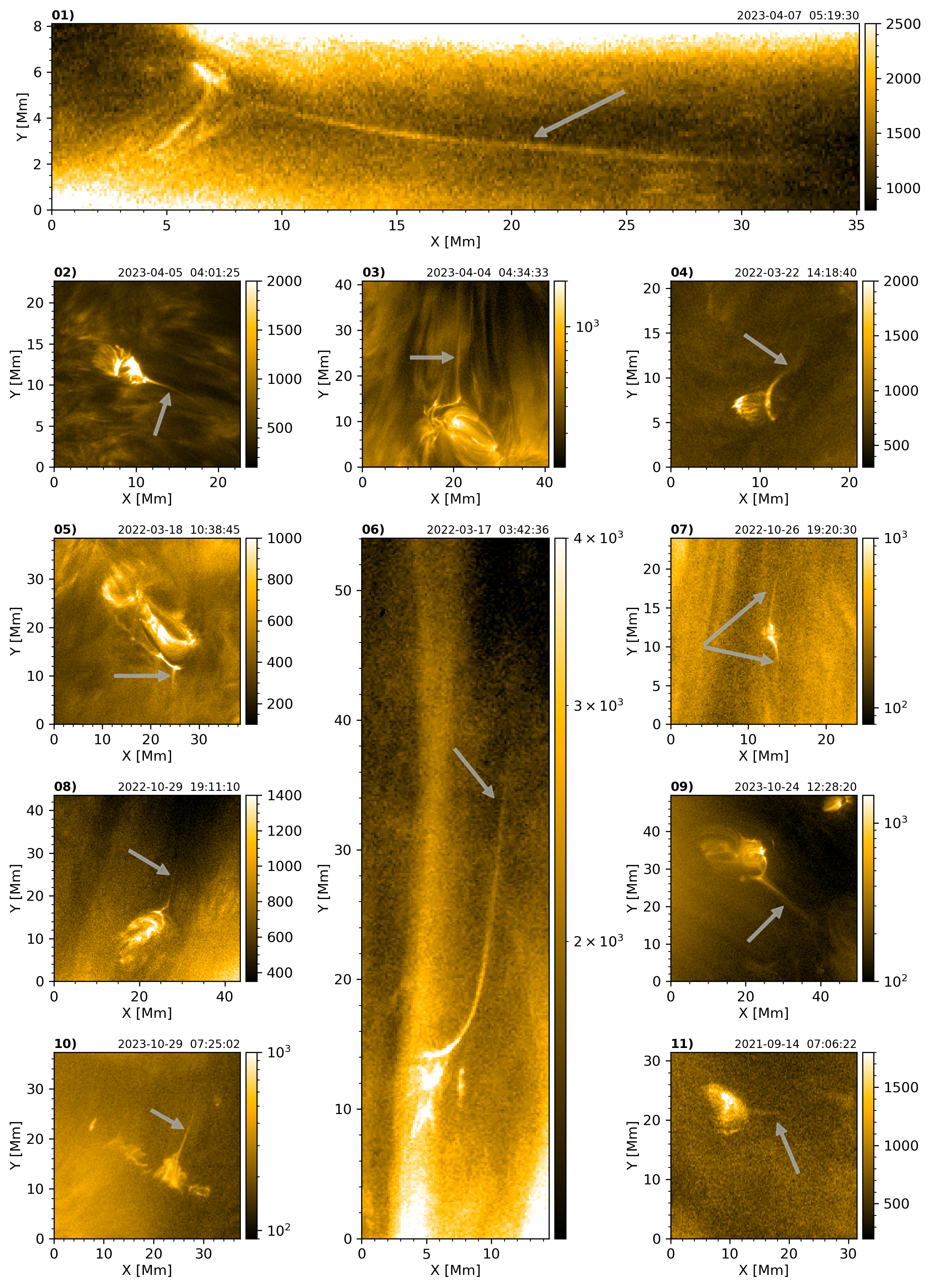}
   \caption{Context for the Solar Orbiter HRI$_{\mathrm{EUV}}$ observations of the eleven cases showing 
   thin coronal jets (gray arrows) associated with CBPs that are analyzed in this paper. 
   The intensity of the images is given in DN and times are in UT.
   The details of each observation are provided in Table~\ref{tab:solo_observations}. 
   Individual movies for each of the cases are available \href{https://zenodo.org/records/16903189}{online}.}
   \label{fig:context}
\end{figure*}

Figure~\ref{fig:context} provides an overview of the eleven HRI$_{\mathrm{EUV}}$ datasets analyzed in this study, showing the CBPs and indicating the location of the associated thin coronal jets with gray arrows.
Since Solar Orbiter moves over time, we present the data using physical X and Y coordinates in megameters, taking into account the varying distance to the Sun as listed in Table~\ref{tab:solo_observations}.
Individual movies for each dataset are available \href{https://zenodo.org/records/16903189}{online}.
For completeness, corresponding SDO/AIA 171~\AA\ and SDO/HMI maps of the regions observed with HRI$_{\mathrm{EUV}}$ are provided in Appendix~\ref{app:sdo} as
Fig.~\ref{fig:appendix_aia} and Fig.~\ref{fig:appendix_hmi}, respectively.
In the following, we briefly describe each case and summarize the properties of the associated jets.
For the reader’s convenience, the projected lengths and widths of the thin jets discussed in this section are later summarized in Table~\ref{tab:jet_properties}.

\begin{figure*}[!ht]
   \centering
   \includegraphics[width=1.0\textwidth]{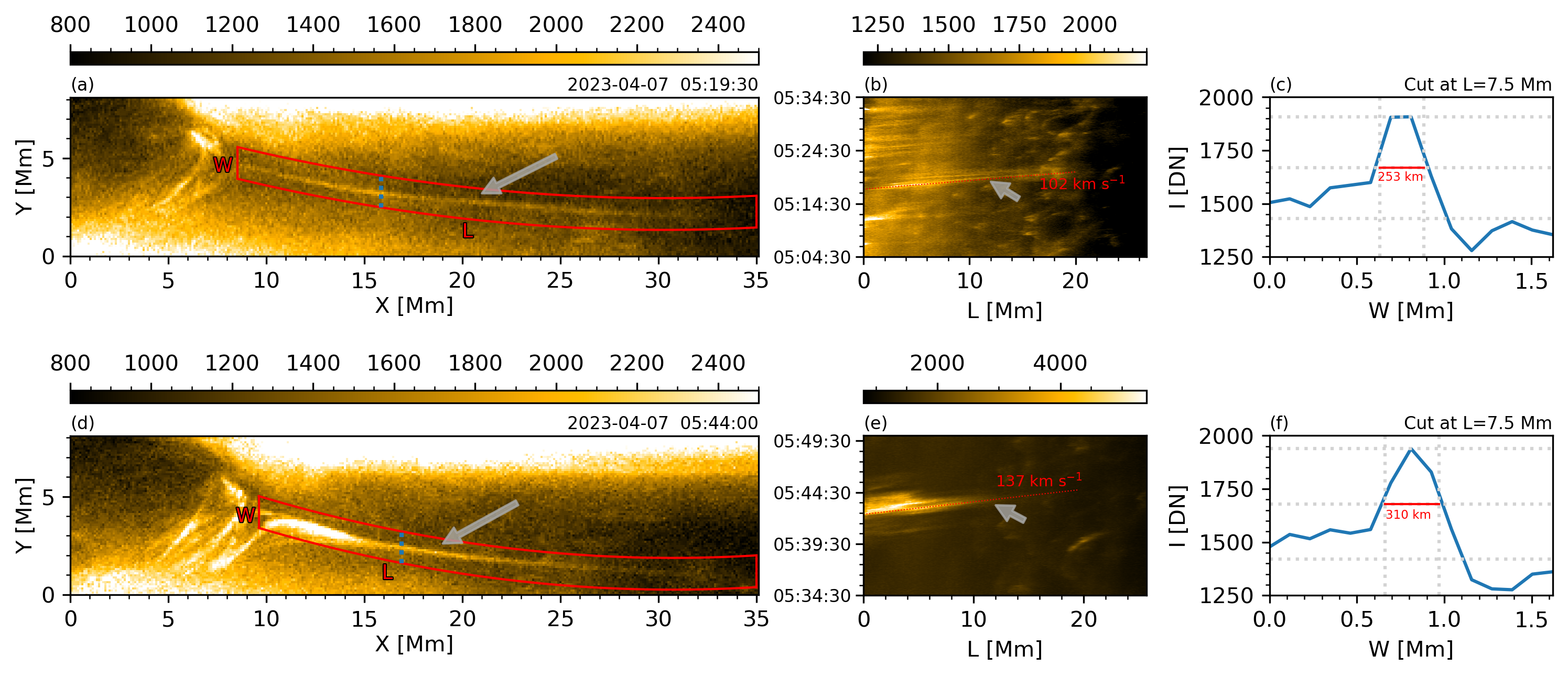}
   \caption{Analysis of Case 01 during the gentle reconnection phase at 05:19:30 UT (top row) and the enhanced phase at 05:44:00 UT (bottom row).
   Panels (a) and (d) show the CBP and the associated thin coronal jets (gray arrows).
   Panels (b) and (e) display space–time maps obtained by taking the maximum intensity along the width $W$ of the bent red slit of length $L$ shown in panels (a) and (d), respectively.
   Panels (c) and (f) illustrate the widths of the narrow jets, calculated using the FWHM at $L = 7.5$~Mm, as indicated by the blue dotted lines in panels (a) and (d), respectively.}
   \label{fig:figure_02}
\end{figure*}

\paragraph{\textit{Case 01: 2023-04-07.}}
This event occurred within a coronal hole and shows two types of behavior: a gentle phase followed by an enhanced phase (see Fig.~\ref{fig:context} and associated animation).
The gentle phase initially exhibited faint brightenings at 04:23:00~UT around coordinates $X = 3.8$ and $Y = 5.5$~Mm.
By 04:33:00~UT, a small inverted Y-shaped structure became apparent, centered at $X = 5.5$ and $Y = 5.6$~Mm.
The activity in this region increased, revealing a distinct fan-spine structure at 04:43:51~UT, with several coronal jets propagating along the outer spine.
At 05:19:30~UT, the longest narrow jet of this phase was observed emerging from the fan base, extending across the entire HRI$_{\mathrm{EUV}}$ field of view (FOV).
In Fig.~\ref{fig:figure_02}, we analyze the properties of this jet.
Panel (a) shows the context, indicating the location of the thin jet with an arrow.
We have also superimposed a bent slit of width $W$ and length $L$ to characterize the jet, following an approach as \citet{JoshiR_etal:2017} and \cite{Petrova_etal:2024}.
Panel (b) presents the corresponding space-time plot, obtained by extracting the maximum intensity along $W$ from the region defined in panel (a).
There, we observe that the jet reaches a projected length of $L_{\mathrm{jet}} = 22.1$~Mm and a velocity of 102 km s$^{-1}$.
In addition, several other jets are visible before and after this one, exhibiting comparable velocities but apparent lengths below 8~Mm.
Panel (c) displays the intensity profile along $W$ at $L = 7.5$~Mm.
The jet width is determined from the full width at half maximum (FWHM) of the intensity profile.
The baseline for the FWHM is set as the average intensity at the endpoints of the perpendicular cut.
The resulting width shows that this long jet is as thin as $w_{\mathrm{jet}}=253$~km.
Given that in Case 01 the pixel size is 108~km (Table~\ref{tab:solo_observations}), this example demonstrates HRI$_{\mathrm{EUV}}$’s ability to resolve very narrow coronal jets.
In fact, this event is not detectable in AIA (see Appendix~\ref{app:sdo}).

The enhanced phase occurs later, between 05:42:00~UT and 05:44:30~UT, and leads to the brightest jet observed in Case 01.
Panel (d) of Fig.~\ref{fig:figure_02} illustrates this jet, which exhibits a much broader base compared to those seen during the gentle phase, and reaches a projected length of $L_{\mathrm{jet}} = 21.7$~Mm.
The speed of the jet is approximately 137 km s$^{-1}$ as shown in panel (e).
By measuring its FWHM at $L = 7.5$~Mm, we find that the jet width is $w_{\mathrm{jet}}=310$~km.
The transition from the gentle to the enhanced phase 
possibly reflects different reconnection regimes.
As discussed later in Sect.~\ref{s:indirect_plasmoids}, this enhanced phase may result from plasmoid-mediated magnetic reconnection.

\paragraph{\textit{Case 02: 2023-04-05.}}
The observation for Case~02 was conducted in two sequences (see Table~\ref{tab:solo_observations}).
According to HMI data (Fig.~\ref{fig:appendix_hmi}), this CBP exhibits a canonical photospheric magnetic field configuration typical of CBPs: a parasitic polarity, negative in this case, surrounded by an oppositely signed magnetic field.
Figure~\ref{fig:context} shows the CBP of interest, centered at $X=8$ and $Y=12.5$~Mm, with a diameter of approximately 4 Mm, placing it at the lower end of the size distribution \citep{Madjarska:2019}.
The CBP brightness and associated jet activity fluctuate over time across both sequences.
The longest coronal jet emanating from the CBP is clearly visible at 04:01:25 UT (the time of the image), with a length of $L_{\mathrm{jet}}=9.1$~Mm and a width of $w_{\mathrm{jet}}=326$~km (see Appendix~\ref{app:width} and Fig.~\ref{fig:appendix_width} for the calculation details).

\paragraph{\textit{Case 03: 2023-04-04.}}
Similar to Case 02, this event is also observed in two sequences.
If inspected using HMI (Fig.~\ref{fig:appendix_hmi}), this area seems to correspond to an emerging region where the parasitic polarity is positive. 
In Fig.~\ref{fig:context}, a well-defined and broad CBP, with an elliptical shape and a major axis diameter of approximately 20 Mm, is observed just above the edge of the HRI$_{\mathrm{EUV}}$ FOV.
A thin jet is visible from the very first frame of observations, departing from the CBP at $X=20$~Mm and $Y=14$~Mm.
It exhibits a recurring pattern throughout the entire duration of the first sequence.
At the time shown in Fig.~\ref{fig:context}, the jet has a length of $L_{\mathrm{jet}}=15.4$~Mm and a width of $w_{\mathrm{jet}}=629$~km (see also Appendix~\ref{app:width}).
A series of broader jets are also observed adjacent to the thin outer spine, appearing at 04:47:58 UT, followed by subsequent occurrences at 05:05:18, 06:01:58, and 06:09:23 UT.
During the second sequence, the long and thin jet from the CBP becomes visible at 06:33:29 UT, followed by the emergence of broad jets originating from the CBP footpoints adjacent to it.

\paragraph{\textit{Case 04: 2022-03-22.}}
The first EUV signatures become visible at 14:06:10~UT, quickly developing the characteristic loops with enhanced EUV emissivity typical of a CBP. 
The CBP seems to be nearby a region exhibiting a clear negative parasitic polarity (see Fig.~\ref{fig:appendix_hmi}).
The maximum projected diameter of the CBP is around 6 Mm (Fig.~\ref{fig:context}), which falls within the lower range of CBP sizes \citep{Madjarska:2019}.
A thin coronal jet with width $w_{\mathrm{jet}}=374$~km is ejected from the top of the CBP, reaching its longest extension $L_{\mathrm{jet}}=4.1$~Mm at 14:23:10 UT. 
Subsequently, the CBP dissipates at 14:25:40 UT, resulting in a CBP lifetime of around 19 minutes.

\paragraph{\textit{Case 05: 2022-03-18.}}
This large CBP remains visible throughout the entire time series with an elliptical shape and a major axis diameter of 20~Mm, approximately.
The jet of interest starts appearing at the corner of the CBP around 10:36:35 UT, specifically, at $X=24$~Mm, $Y=12$~Mm, following a series of complex dynamic events.
We could tentatively relate this event to a dark, elongated structure that might correspond to a small-scale filament, consistent with those reported by, for instance, \cite{Panesar_etal:2018} and \cite{Kumar_etal:2019a}, or to a rising chromospheric fibril, as seen in the observations by \cite{Madjarska_etal:2021} and in the 3D simulations of \cite{Nobrega-Siverio_etal:2023}.
However, in the absence of coordinated chromospheric observations, we refrain from drawing firm conclusions.
The current sheet associated with this jet is also clearly discernible, with a length of $L_{\mathrm{CS}}=2.8$~Mm, approximately.
At the time shown in Fig.~\ref{fig:context}, the jet shows a nice contrast against the background and has a length of $L_{\mathrm{jet}}=5.8$~Mm and a width of $w_{\mathrm{jet}}=481$~km (see Appendix~\ref{app:width}).
Later, at 11:04:55 UT, a dark, absorbing structure interacts with the current sheet, triggering a more eruptive event with a broader jet.

\paragraph{\textit{Case 06: 2022-03-17.}}
The 30-minute sequence reveals a rapidly evolving CBP with a maximum projected diameter of approximately 6 Mm, also placing this event at the lower end of the CBP size distribution \citep{Madjarska:2019}.
HMI data indicate that this case corresponds to a CBP with a negative parasitic polarity in the photosphere (Fig.~\ref{fig:appendix_hmi}).
The largest thin coronal jet associated with this CBP occurs at 03:42:36~UT (Fig.~\ref{fig:context}) with a length of $L_{\mathrm{jet}}=21.8$~Mm and a width of $w_{\mathrm{jet}}=344$~km (see Appendix~\ref{app:width}), and has a counterpart in AIA 171~\AA\ (see Fig.~\ref{fig:appendix_aia}).

\paragraph{\textit{Case 07: 2022-10-26.}}
Case 07 corresponds to a small and short-lived CBP.
The first signatures of the CBP appear around 19:14:20 UT.
HMI data suggest that an emerging dipole, with a positive parasitic polarity, may have occurred at this location.
By 19:17:45 UT, a distinct bidirectional ejection emerges from the CBP, featuring two spines extending to the north and south.
The bidirectional jets reach their maximum extent at 19:20:30 UT,
with lengths around $L_{\mathrm{jet}}=5.5$~Mm and widths of $w_{\mathrm{jet}}=382$~km (see Appendix~\ref{app:width}).
In AIA 171~\AA, one might infer their presence based on the HRI$_{\mathrm{EUV}}$ observations (Fig.~\ref{fig:appendix_aia}).
The bidirectional jet configuration resembles that studied in AIA by \citet{Ruan:etal:2019} and in numerical experiments with horizontal coronal magnetic field \citep[e.g.,][]{Yokoyama_Shibata:1996,Archontis_etal:2005,Syntelis_etal:2019, Faerder_etal:2023}.
In fact, just before the jets, one may identify a dark structure ejected above the CBP, which could represent part of the emerging dome being expelled as a plasmoid or surge, as shown in simulations \citep[see][among others]{Archontis_etal:2006,Nishizuka_etal:2008,Jiang_etal:2012,
Moreno-Insertis_Galsgaard:2013,Nobrega-Siverio_etal:2016,Nobrega-Siverio_etal:2018}, although
chromospheric observations are required for a clear conclusion.
Afterward, the CBP gradually fades and disappears by 19:22:05 UT, resulting in a total lifetime of approximately 8 minutes.

\paragraph{\textit{Case 08: 2022-10-29.}}
This CBP, akin to Cases 02 and 03, was observed in two sequences, although there is a 2.5-hour gap between them and the HRI$_{\mathrm{EUV}}$ exposure time was adjusted for the second one (see Table~\ref{tab:solo_observations}).
The CBP, with a projected diameter of approximately 15~Mm, is already visible in the first frame of the first sequence and remains discernible until the end of the second sequence, resulting in a lifetime of more than 3.5 hours.
HMI data reveal that the CBP of interest is rooted in a negative parasitic polarity, surrounded by a region of positive magnetic polarity.
The animation shows recurrent coronal jets that are faint and narrow. 
For instance, the jet shown in Fig.~\ref{fig:context} reaches a length of $L_{\mathrm{jet}}=19.5$~Mm and a width of $w_{\mathrm{jet}}=411$~km (see also Appendix~\ref{app:width}).
All these jets originate from a very thin and highly dynamic current sheet located at the top of the CBP, centered at $X = 26$~Mm and $Y = 17$~Mm.
This current sheet is particularly interesting, as it shows clear signatures of non-stationary magnetic reconnection, including the presence of a well-defined plasmoid.
A detailed analysis of the plasmoid and its dynamic behavior will be presented in Sect.~\ref{s:direct_plasmoids}.

\paragraph{\textit{Case 09: 2023-10-24.}}
This event was observed for nearly 2 hours.
At the beginning of the HRI$_{\mathrm{EUV}}$ sequence, the CBP exhibits an ellipse-like shape with a major axis of approximately 25~Mm.
As in other cases, HMI shows that the CBP is rooted in a negative parasitic polarity.
The CBP is highly dynamic and displays a microflare-like structure around 12:24:00 UT, accompanied by a collimated flow.
Similar to Case 05, this event may be related to enhanced chromospheric activity, possibly involving either a small-scale filament or a rising fibril; 
nonetheless, coordinated chromospheric observations are required to disentangle the scenario.
The thin jet observed at 12:28:20 UT in Fig.~\ref{fig:context} reaches a length of $L_{\mathrm{jet}}=16.1$~Mm and a width of $w_{\mathrm{jet}}=706$~km, and disappears shortly afterward (see Appendix~\ref{app:width}). 
In AIA 171~\AA, it is also possible to discern it (Fig.~\ref{fig:appendix_aia}).
An analogous jetting pattern is seen at 12:32:40, 12:58:30, 13:15:50, and 13:44:50 UT. 
By the end of the observational sequence, the CBP had become narrower compared to its initial state.

\paragraph{\textit{Case 10: 2023-10-29.}}
In this case, the HRI$_{\mathrm{EUV}}$ FOV encompasses multiple adjacent CBPs, the largest of which is located at approximately $X = 23$ and $Y = 13$~Mm.
HMI magnetograms reveal three negative polarity patches embedded within a dominant positive field, which may account for the presence of multiple CBPs in the region.
The largest CBP has a very distinguishable fan base with a thin jet extending from its top at 07:15:32~UT. 
In line with other events, the jetting activity is intermittent. 
At 07:25:02~UT (see Fig.~\ref{fig:context}), the CBP displayed the clearest narrow jet from its base, with a length of $L_{\mathrm{jet}}=7.5$~Mm and a width of $w_{\mathrm{jet}}=637$~km, which disappeared by the end of the HRI$_{\mathrm{EUV}}$ observation at 07:29:52~UT (see also Appendix~\ref{app:width}).

\paragraph{\textit{Case 11: 2021-09-14.}}
Among the events studied here, this CBP is captured from the greatest distance and with the shortest exposure time (see Table~\ref{tab:solo_observations}), which explains why the image quality appears slightly lower than in the other cases.
The CBP is present throughout the entire time series with a diameter of 6~Mm, approximately, and it is anchored in a region with a positive parasitic polarity, as seen in HMI.
A very thin jet appears at 07:05:02~UT and remains discernible for almost 6 minutes. 
At the time shown in Fig.~\ref{fig:context}, the jet has an extension of $L_{\mathrm{jet}}=6.7$~Mm and a width of $w_{\mathrm{jet}}=591$~km.
Due to the short duration of this observation, the reappearance of the outer spine, as seen in most other cases, could not be captured.

\begin{table}[htbp]
\centering
\caption{
 Properties of the CBPs and associated thin coronal jets.}
\begin{tabular}{cccc}
\hline\hline
\rule{0pt}{2.5ex}
Case & $D_\mathrm{CBP}$ (Mm) & $L_{\mathrm{jet}}$ (Mm) & $w_{\mathrm{jet}}$ (km) \\
\hline
\rule{0pt}{2.5ex}
01 & 5  & 22.1 & 253 \\
   &    & 21.7 & 310 \\
\rule{0pt}{1.5ex}
02 & 4  & 9.1  & 326 \\
\rule{0pt}{1.5ex}
03 & 20 & 15.4 & 629 \\
\rule{0pt}{1.5ex}
04 & 6  & 4.1  & 374 \\
\rule{0pt}{1.5ex}
05 & 20 & 10.0 & 481 \\
\rule{0pt}{1.5ex}
06 & 6  & 21.8 & 344 \\
\rule{0pt}{1.5ex}
07 & 3  & 5.5  & 382 \\
\rule{0pt}{1.5ex}
08 & 15 & 19.5 & 411 \\
\rule{0pt}{1.5ex}
09 & 25 & 16.1 & 706 \\
\rule{0pt}{1.5ex}
10 & 6  & 7.5  & 637 \\
\rule{0pt}{1.5ex}
11 & 6  & 6.7  & 591 \\
\hline
\end{tabular}
\tablefoot{$D_\mathrm{CBP}$, projected diameter of the CBP. $L_{\mathrm{jet}}$ and  $w_{\mathrm{jet}}$, projected length and width, respectively, of the associated thin coronal jets shown in Fig.~\ref{fig:context}, Fig.~\ref{fig:figure_02}, and Fig.~\ref{fig:appendix_width} of this study.}
\label{tab:jet_properties}
\end{table}

\section{Signatures of plasmoid-mediated reconnection: comparison with numerical models}\label{s:results_reconnection}

\subsection{Direct plasmoid detection}\label{s:direct_plasmoids}
\begin{figure*}[!ht]
   \centering
   \includegraphics[width=1.0\textwidth]{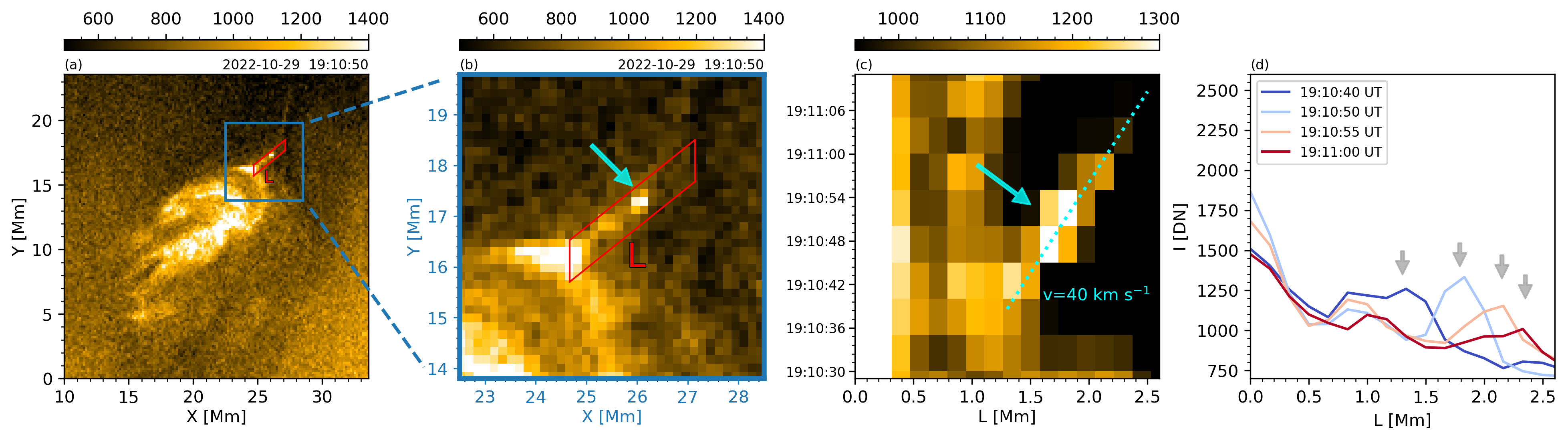}
   \caption{Plasmoid signatures in Case 08. 
   Panel (a): Context view showing the CBP and the current sheet within a rectangle of length $L$ and width $W$.
   Panel (b): Zoomed-in view of the blue rectangle shown in panel (a), highlighting the illustrative plasmoid with a cyan arrow.
   Panel (c): Space–time plot of the current sheet, obtained by taking the maximum intensity along the $W$ direction. 
   The cyan dashed line indicates the trajectory of the plasmoid.
   Panel (d): Intensity profiles along the current sheet at different times, illustrating the evolution of the plasmoid indicated in panel (c).
   An animation of this figure is available \href{https://zenodo.org/records/16903189}{online} showing the evolution of the plasmoid between 19:10:35 UT and 19:11:10 UT.
   }
   \label{fig:figure_03}
\end{figure*}

We have detected a clear case of a plasmoid being formed and ejected within the current sheet of the CBP in Case 08.
To illustrate this, we refer to Fig.~\ref{fig:figure_03} and the accompanying animation.
Panel~(a) shows the CBP and the associated current sheet, highlighted by a red rectangle of length $L$ and width $W$, 
at the time when the plasmoid is most clearly visible (19:10:50 UT).
Panel~(b) presents a zoom into the region around the current sheet, corresponding to the blue rectangle in panel (a).
The cyan arrow indicates a bright, blob-like feature, which we interpret as a plasmoid. 
Its size is approximately two pixels in each direction, corresponding to about 332 km.
To analyze its motion, we measured the maximum intensity along the $W$ direction within the slanted red rectangle. 
The results are shown as a space–time plot in panel (c).
The plasmoid displays a projected velocity of approximately 40~km s$^{-1}$.
Panel (d) contains intensity profiles along the current sheet at different times, tracking the temporal evolution of 
the detected plasmoid.
From this panel and the accompanying animation, we estimate the plasmoid's lifetime to be around 20 seconds.

\begin{figure*}[!ht]
   \centering
   \includegraphics[width=1.0\textwidth]{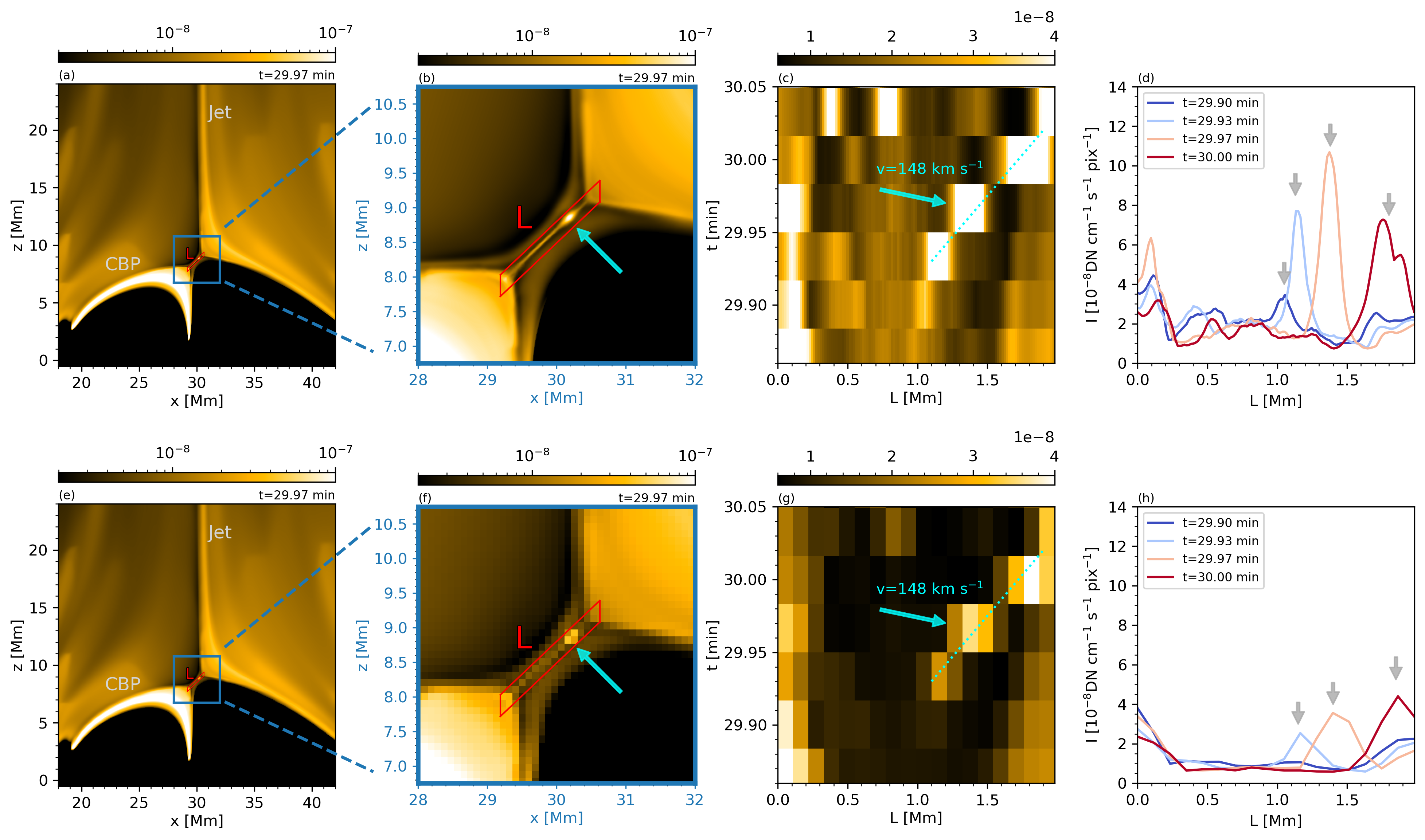}
   \caption{Same as Figure~\ref{fig:figure_03}, but for the plasmoid signatures in the 2D CBP simulation by \cite{Nobrega-Siverio_Moreno-Insertis:2022}.
   The top row shows the HRI$_{\mathrm{EUV}}$ synthetic emissivity at the original resolution of the numerical experiment ($\approx 16$~km).
   The bottom row shows the results after degrading the resolution to match the best case achievable by HRI$_{\mathrm{EUV}}$ in our study (pixel size of 108~km).
   An animation of this figure is available \href{https://zenodo.org/records/16903189}{online} showcasing the plasmoid evolution between $t=29.9$ and $t=30.0$ min. 
   }
   \label{fig:figure_04}
\end{figure*}

To support this finding, we compare the Solar Orbiter results with those from the CBP numerical experiment (see Sect.~\ref{s:methods_sim}).
Figure~\ref{fig:figure_04} and the associated animation show the synthetic HRI$_{\mathrm{EUV}}$ emissivity derived from 
the simulation, using the same panel layout as in Fig.~\ref{fig:figure_03}.
The top row displays the simulation at its native resolution ($\approx16$~km), while the bottom row 
shows the same results after degrading the image to the best-case scenario achieved by HRI$_{\mathrm{EUV}}$ in this study (pixel size of 108~km).
The first column provides context at a time when a clear plasmoid is present ($t = 29.97$~min), showing enhanced synthetic emissivity in the CBP loops (located between $x = 19$ and $30$~Mm) and the jet spine extending upward from $z = 9$~Mm 
at $x = 30.4$~Mm.
In the second column, a bright plasmoid can be clearly distinguished within the current sheet.
At the native resolution (panel b), the plasmoid has an estimated length of 300~km and a width of 200~km, approximately,
which is a size large enough to be resolved by HRI$_{\mathrm{EUV}}$.
Panel (f) confirms this: although the plasmoid appears more subtle after resolution degradation, it remains detectable 
as a cluster of pixels with enhanced emissivity, in close resemblance to the observational case.
The third column shows the plasmoid's motion through space–time maps. 
Its signature resembles the observational one shown in Fig.~\ref{fig:figure_03}, although it travels faster in 
the simulation, with a velocity of 148~km~s$^{-1}$ (see Sect.~\ref{s:discussion} for a discussion on the differences in plasmoid velocities).
Finally, the fourth column displays the intensity profiles along the current sheet at different times, capturing 
the temporal evolution of the plasmoid.
Even after degrading the resolution, the plasmoid remains detectable through triangular-shaped intensity peaks moving 
along the current sheet, consistent with the observational findings.

\subsection{Indirect signatures of plasmoids}\label{s:indirect_plasmoids}

\begin{figure*}[!ht]
   \centering
   \includegraphics[width=1.0\textwidth]{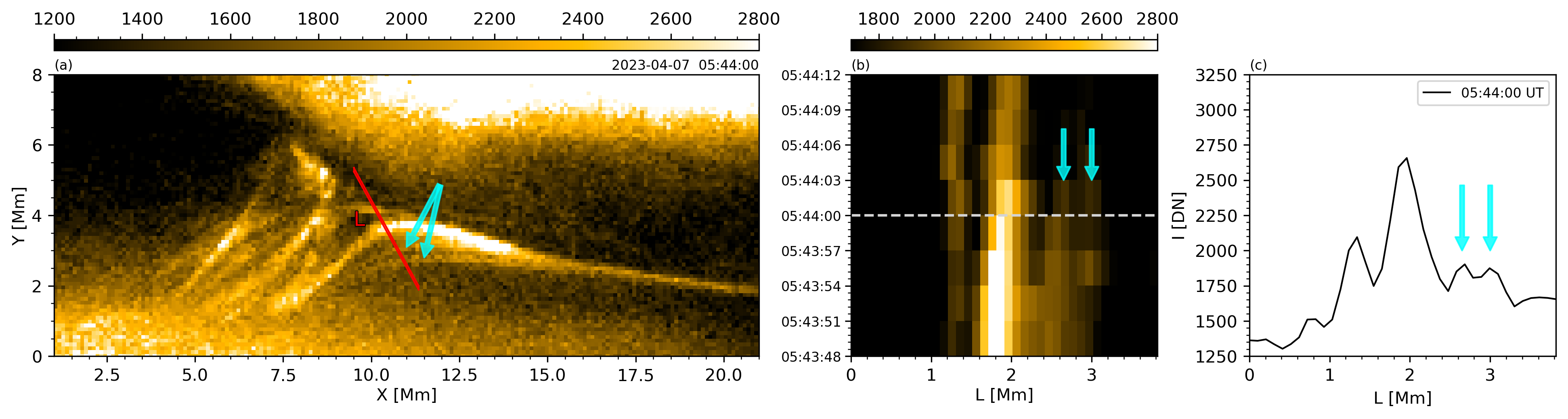}
   \caption{Indirect signatures of plasmoid-mediated reconnection in Case 01.
   Panel (a): Context view illustrating the CBP and the associated ejections. 
   The arrows mark the signatures that are interpreted as consequences of plasmoid-mediated magnetic reconnection.
   Panel (b): Space–time map obtained by sampling the intensity along the slit of length $L$ shown in panel (a).
   Panel (c): Intensity profiles along the slit at 05:44:00 UT.
   An animation of this figure is available \href{https://zenodo.org/records/16903189}{online}, showing the time evolution between 05:43:48 UT and 05:44:12 UT.
   }
   \label{fig:figure_05}
\end{figure*}

In Case 01, we find imprints in the outflow region that could be a consequence of plasmoid-mediated reconnection.
Since no plasmoids are directly detected in HRI$_{\mathrm{EUV}}$, we refer to these as potential indirect signatures.
This is illustrated in Fig.~\ref{fig:figure_05} and the associated animation.
In panel (a), the brightest jet is shown to provide context, with arrows indicating a double adjacent boomerang-like discontinuous pattern that appears slightly brighter than its surroundings.
To further inspect this feature, in panel (b) we present a space-time map of the intensity along the slit of length $L$ indicated in panel (a).
The structure appears as two nearly vertical bands, with an apparent velocity of only a few kilometers per second.
In panel (c), we show the intensity profile along the slit at time 05:44:00~UT.
The arrows indicate two small humps, standing slightly enhanced compared to the surrounding regions.
We conjecture that this feature is a consequence of plasmoids being ejected and colliding with the pre-existing magnetic field, potentially explaining the transition from a gentle reconnection phase to an enhanced phase associated with the brightest jet of Case 01.

\begin{figure*}[!ht]
   \centering
   \includegraphics[width=1.00\textwidth]{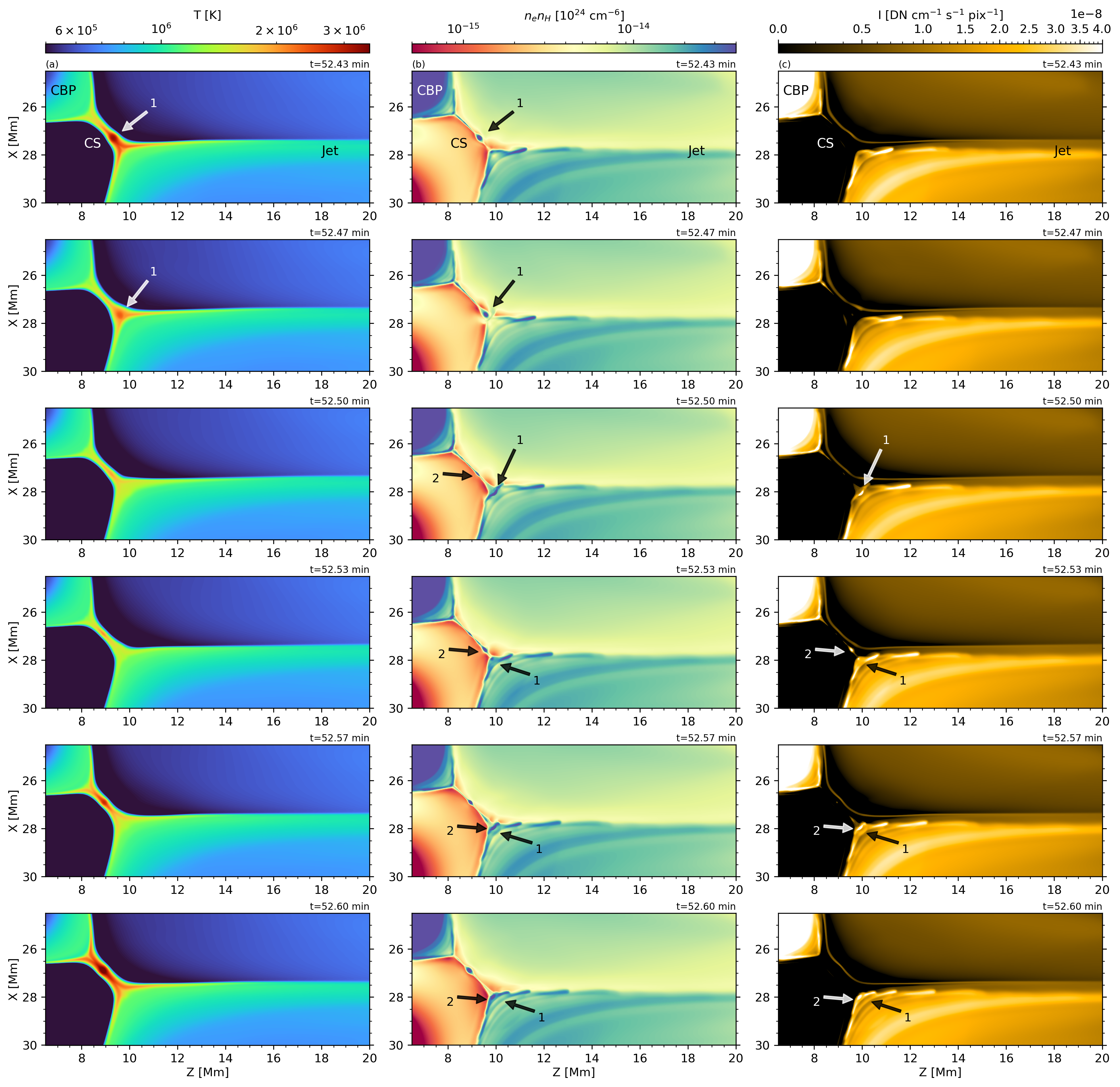}
   \caption{Indirect signatures of plasmoid-mediated reconnection in the outflow region from the 2D CBP simulation by \cite{Nobrega-Siverio_Moreno-Insertis:2022} between $t=52.43$ and $t=52.60$ min.
   We have transposed the axes so that height appears on the horizontal axis to ease the comparison with the observations.
   Column (a): Temperature maps.
   Column (b): $n_e n_H$, the product of the electron and hydrogen number densities.
   Column (c): HRI$_{\mathrm{EUV}}$ synthetic emissivity at the original resolution of the experiment ($\approx16$~km).
   The three panels of the top row include the locations of the CBP, the current sheet (CS), and the jet.
   The arrows 1 and 2 follow the formation and ejection of two different plasmoids along with their signatures in the outflow region.
   }
   \label{fig:figure_06}
\end{figure*}

\begin{figure*}[!ht]
   \centering
   \includegraphics[width=1.00\textwidth]{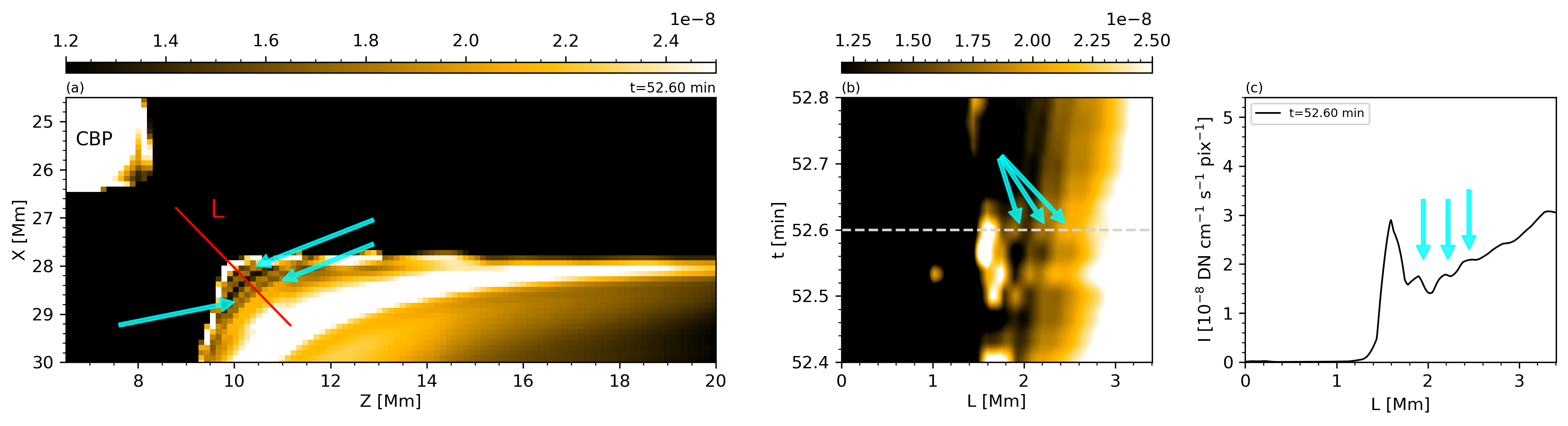}
   \caption{Indirect signatures of plasmoid-mediated reconnection in the outflow region from the 2D CBP simulation by \cite{Nobrega-Siverio_Moreno-Insertis:2022} after degrading the resolution to match the best case achievable by HRI$_{\mathrm{EUV}}$ in our study (pixel size of 108~km).
   Panel (a): HRI$_{\mathrm{EUV}}$ synthetic emissivity.
   We have transposed the axes so that height appears on the horizontal axis.
   Panel (b): Space–time map obtained by sampling the intensity along the slit of length $L$ shown in panel (a).
   Panel (c): Intensity profiles along the slit at $t=52.60$~min.
   In all the panels, the arrows mark some of the features that are interpreted as consequences of plasmoids previously ejected during the reconnection process.
   An animation of this figure is available \href{https://zenodo.org/records/16903189}{online}, including temperature and $n_e n_H$ maps,  showcasing the appearance of indirect plasmoid-mediated signatures between $t=52.40$ and $t=52.80$ min.} 
   \label{fig:figure_07}
\end{figure*}

To support this hypothesis, we turn to simulations, from an instant approximately 22~min later than that shown in Fig.~\ref{fig:figure_04}. 
Figure~\ref{fig:figure_06} shows maps over a 10-second time window of the temperature (column a); $n_e n_H$, the product of the electron and hydrogen number densities, which is a key factor in computing the emissivity under optically thin conditions (column b); and the synthetic HRI$_{\mathrm{EUV}}$ emissivity at the simulation's native resolution (column c).
In all maps, the axes have been transposed so that height appears along the horizontal axis, mimicking the morphology of the jet observed in Case 01.
The top-row maps highlight the location of the post-reconnection loops that give rise to the CBP, the jet spine, and the current sheet (CS).
A plasmoid, labeled as 1, is identified within the current sheet in both the temperature and $n_e n_H$ maps between $t=52.43$~min and $t=52.47$~min. 
However due to its high temperature ($\approx 3$~MK), it is not directly visible in HRI$_{\mathrm{EUV}}$, whose 
response function peaks at $\approx 0.9$~MK \citep[see, e.g.,][]{Chen_etal:2021,Dolliou_etal:2023}.
At $t=52.50$~min, plasmoid 1 collides with the pre-existing field, generating a shock and losing its identity as a closed magnetic island via reconnection.
Consequently, the density originally confined within the plasmoid is redistributed along the newly reconnected field line.
Following the subsequent evolution in the outflow region, a boomerang-like structure develops, exhibiting slightly enhanced density and HRI$_{\mathrm{EUV}}$ emissivity relative to its surroundings, most clearly seen at $t=52.60$~min.
In this 10-second time window, a second plasmoid, labeled as 2, is also detected within the current sheet at $t=52.50$~min in the $n_e n_H$ map propagating in the same direction as plasmoid~1.
Its increasing density, combined with a lower temperature than that of plasmoid 1, makes it visible two seconds later in the HRI$_{\mathrm{EUV}}$ map ($t=52.52$~min).
Then, it collides with the pre-existing field, generating another boomerang-like structure with enhanced emissivity.
Due to the intermittent nature of plasmoid-mediated reconnection, multiple boomerang-like patterns appear over time in the outflow region of the simulation.

Why do we think this scenario is consistent with the observations?
The simulations show that current sheets can host plasmoids that are challenging to detect with Solar Orbiter HRI$_{\mathrm{EUV}}$. 
Plasmoid 1 is too hot to fall within the instrument’s response range. 
Plasmoid 2, in contrast, evolves so rapidly that it appears in HRI$_{\mathrm{EUV}}$ emissivity only in a single snapshot of the numerical experiment, despite the simulation having a 2-second cadence: a cadence that is already shorter than that of any of the HRI$_{\mathrm{EUV}}$ observations presented in this paper.
Nevertheless, the impact of these plasmoids can still leave observable imprints after they leave the current sheet. 
In particular, the simulations reveal the formation of boomerang-like structures in the outflow region that closely resemble those seen in Case 01.
To further illustrate the similarity with observations, we refer to Fig.~\ref{fig:figure_07} and the associated animation.
Panel (a) shows the emissivity map from the simulation at $t=52.60$~min, degraded to match the best spatial resolution achievable by Solar Orbiter HRI$_{\mathrm{EUV}}$.
Even after degradation, the boomerang-like pattern generated in the outflow region by multiple plasmoids remains discernible (see arrows).
In panel (b), the space–time map, calculated along the slit of length $L$ shown in panel (a), illustrates how the discontinuous pattern propagates at a velocity of approximately $23$~km s$^{-1}$.
In the intensity profile at $t=52.60$~min (panel c), we identify the corresponding small humps, slightly enhanced with respect to the surrounding regions, closely resembling those seen in the observations.
This supports the interpretation that indirect signatures of plasmoid-mediated reconnection in the outflow region may indeed be captured by HRI$_{\mathrm{EUV}}$.

\section{Discussion and conclusions}\label{s:discussion}
In this paper, we have used high-resolution Solar Orbiter HRI$_{\mathrm{EUV}}$ observations to study eleven datasets that show narrow coronal jets associated with CBPs.
In parallel, we have analyzed observational signatures indicative of plasmoid-mediated reconnection, and interpreted them in light of the radiative-MHD numerical simulations published by \cite{Nobrega-Siverio_Moreno-Insertis:2022}.
Importantly, these features have been detected without applying any image enhancement techniques such as 
Multiscale Gaussian Normalization \citep[MGN;][]{Morgan_Druckmuller:2014} or Wavelet-Optimized Whitening
\citep[WOW;][]{Auchere_etal:2023}.
This underscores the state-of-the-art capability of HRI$_{\mathrm{EUV}}$ for capturing fine-scale jets and key dynamic signatures from a vantage point close to the Sun.
In the following, we discuss the broader implications of our findings and present the main conclusions.

\subsection{Thin coronal jets from CBPs}
In Sect.~\ref{s:results_jets}, we have identified coronal jets originating from CBPs with widths ranging from 253~km to 706~km: scales that could not be resolved with previous EUV imaging instruments, where jet widths were typically reported to be above 2~Mm \citep[see, e.g.,][]{Shimojo_etal:1996,Savcheva_etal:2007,Raouafi_etal:2016,JoshiR_etal:2017,Shen:2021,Panesar_etal:2023}.
Remarkably, these jets extend up to projected lengths of 22~Mm (Cases 01 and 06), while maintaining their narrow structure.
This demonstrates that jet outflows can propagate over long distances along narrow structures, and potentially contribute to the solar wind.
The most comparable event observed in HRI$_{\mathrm{EUV}}$ to our longest thin CBP jet (Case 01) is reported by \cite{Petrova_etal:2024}, where the authors detected torsional motions propagating along the outer spine.
Interestingly, these narrow jets can originate from quite compact CBPs. 
In fact, seven out of our eleven CBPs have projected diameters ranging from 6~Mm down to 3~Mm (see Table~\ref{tab:jet_properties}), placing them at the lower end of the CBP size distribution \citep{Madjarska:2019}.
Similar compact bright points associated with narrow jets have also been reported using AIA, near the instrument’s resolution limit \citep{Kumar_etal:2019a,Kumar_etal:2022}.

To place our results in the context of recent HRI$_{\mathrm{EUV}}$ findings, we compare them with the recently reported picoflare jets by \cite{Chitta_etal:2023,Chitta_etal:2025}. 
The picoflare jets are inverted Y-shaped structures that extend only a few hundred kilometers, appearing 10\% to 30\% brighter than the surrounding coronal hole, with lifetimes ranging from 20 to 100 seconds.
Picoflare jets have widths of a few hundred kilometers, comparable to our CBP-associated jets, although the latter are larger in extent and exhibit brightness enhancements between 30\% and 85\% relative to their surroundings (see Appendix~\ref{fig:appendix_width}).
Both types of jets, picoflare jets and the longer CBP-associated thin jets presented here, underscore the capabilities provided by HRI$_{\mathrm{EUV}}$ for studying fine-scale coronal dynamics.

In the HRI$_{\mathrm{EUV}}$ time windows analyzed for the eleven cases, we do not observe large blowout jets and associated filament eruptions in line to what have been studied in some numerical models \cite[e.g.,][among others]{Archontis_Hood:2013,Moreno-Insertis_Galsgaard:2013,Fang_etal:2014,Pariat_etal:2015,Lee_etal:2015,Wyper_etal:2017,Wyper_etal:2018a,Wyper_etal:2018b,JoshiR_etal:2024a,Zhuleku_etal:2025,Patsourakos_Archontis:2025}.
Only two of the cases may be tentatively related to some enhanced chromospheric activity in the form of small-scale filament or chromospheric fibril; nonetheless, since no chromospheric observations are available for these events, a conclusive interpretation cannot be established.
Thus, the majority of our observed thin coronal jets seem to be related to the more “gentle” phases of CBP evolution.
These stages are characterized by narrow jets
of the kind of straight jets modeled by, for instance, \citet{Pariat_etal:2010,Pariat_etal:2015}, which can be produced through an ad-hoc photospheric driving and are not impulsively generated, unlike helical or blowout jets.
More recent radiative-MHD models building upon this framework, such as \citet{Nobrega-Siverio_Moreno-Insertis:2022} in 2D and \cite{Nobrega-Siverio_etal:2023} in 3D, have demonstrated that stochastic granular motions are sufficient to stress the CBP’s fan-spine configuration and trigger sustained reconnection at the coronal null point. 
This reconnection naturally results in persistent jetting activity in the form of thin, recurrent jets.

What are the keys to detecting narrow coronal jets associated with CBPs?
High spatial resolution from HRI$_{\mathrm{EUV}}$ appears to be essential to resolve the outer spine of the jets, as evidenced by the comparison presented in Appendix~\ref{fig:appendix_aia}, where some of the jets observed with HRI$_{\mathrm{EUV}}$ are barely discernible or completely absent in AIA~171~\AA. 
%
This is particularly exciting for the coronal jet community. 
For instance, using AIA, \cite{Kumar_etal:2019a} explicitly note that CBPs do not produce coronal jets immediately after emergence, reporting delays ranging from approximately two hours for small CBPs to five days for the largest cases in their sample 
(see also the review by \citealp{Madjarska:2019}). 
\cite{Muglach:2021} warned that “a lack of a jet in AIA during flux emergence should be considered with caution though, as \cite{Young:2015} has shown that there can be coronal hole jets without visible signatures in AIA images.” 
Thanks to the enhanced spatial resolution of HRI$_{\mathrm{EUV}}$, we may now detect earlier or fainter phases of jet activity related to CBPs.

Alongside spatial resolution, the density contrast between the jet and the surrounding atmosphere appears to be the other decisive factor.
Indeed, CBPs can produce collimated high-speed hot plasma outflows that do not exhibit significant EUV emission due to a relatively low density contrast with their surroundings, as demonstrated in the synthetic observables from the 3D CBP numerical experiment of \cite{Nobrega-Siverio_etal:2023}.
In fact, there is observational evidence of CBP jets being detected in EUV spectroscopic data from Hinode/EIS, while lacking any corresponding signal in AIA imaging \citep{Young:2015}, as well as studies aiming to understand the faintness of solar coronal jets from CBPs located in coronal holes \citep{Harden_etal:2021}.
This highlights the need not only for high-resolution imaging, but also spectroscopic diagnostics to fully capture the diversity and subtlety of jet activity associated with CBPs.
Upcoming missions like the Multi-slit Solar Explorer \citep[MUSE;][]{Cheung_etal:2022,De-Pontieu_etal:2022} and Solar-C/EUVST \citep{Shimizu_etal:2020} are expected to significantly advance this goal by delivering coordinated multi-wavelength observations with unprecedented spatial and temporal coverage.

Beyond the challenges of detection, an important open question is whether the jets emerging from CBPs can contribute to the solar wind and its variability.
Current observational efforts are combining Parker Solar Probe \citep[PSP;][]{Fox_etal:2016} and AIA data to assess whether jets from CBPs can be precursors of magnetic switchbacks \citep[e.g.,][]{Kumar_etal:2023,HouC_etal:2024,
Bizien_etal:2025}.
At the same time, numerical models suggest that interchange reconnection in the low corona, particularly in fan-spine magnetic topologies akin to those of CBPs, can produce jets and Alfvénic perturbations capable of propagating into the heliosphere \citep[see, e.g.,][]{Wyper_etal:2022,Bale_etal:2023,Touresse_etal:2024}.
Our detection of narrow jets with HRI$_{\mathrm{EUV}}$ reveals a richer and more dynamic jetting activity from CBPs than what was previously accessible, suggesting that CBPs might play a more significant role in shaping the solar wind.
This opens up exciting prospects for future studies aiming to clarify the connection between jets from CBPs and in situ solar wind structures.

\subsection{Signatures of plasmoid-mediated magnetic reconnection}

\subsubsection{Direct plasmoid detection}
In Case 08, the CBP exhibits a thin current sheet of approximately 2.6~Mm in length, from which coronal jets originate.
Within this current sheet, we identify a clearly resolved plasmoid forming, reaching a size of 332~km, and propagating at an approximate velocity of 40~km~s$^{-1}$ (Sect.~\ref{s:direct_plasmoids}).
The estimated lifetime of the plasmoid is around 20 seconds.
The simulation by \cite{Nobrega-Siverio_Moreno-Insertis:2022} shows that the null point of a fan-spine magnetic topology can be self-consistently stressed by perturbations driven by granular motions.
This leads to the formation of a current sheet with a length comparable to that inferred from the observations, within which a plasmoid develops and reaches a size similar to the observed one. 
When the synthetic emission is degraded to HRI$_{\mathrm{EUV}}$ resolution, the resulting imprints closely match those observed, despite the simulated plasmoid exhibiting a higher velocity of 148 km s$^{-1}$.
This similarity is striking, considering that this numerical experiment is not intended as a direct reproduction of the observed event, but rather as a proof of concept for the observable signatures of plasmoid-mediated reconnection.
The difference in velocity may arise from projection effects; a higher reconnection rate in the simulation, since plasmoid speeds scale with the Alfvén speed  \citep[e.g.,][]{Nishida_etal:2009,Nishizuka_etal:2015}; different plasma beta \citep[][]{Peter_etal:2019}; among others.
Notably, our observational result also aligns with the recent simulations by \cite{Faerder_etal:2024b}, who reported plasmoids with sizes between 200–500~km, lifetimes of 10–20s, and velocities up to 50~km~s$^{-1}$ in coronal fan-spine topologies, concluding that such features should be observable with HRI$_{\mathrm{EUV}}$.

Plasmoids have a long and rich history in physics \citep[][]{Furth_etal:1963}, and their theoretical understanding has significantly advanced, with key developments in the regime of fast reconnection driven by plasmoid instabilities \citep[e.g.,][among many others]{Loureiro_etal:2007,Bhattacharjee_etal:2009,Uzdensky_etal:2010,Loureiro_etal:2012}.
To our knowledge, the direct evidence presented here constitutes the smallest plasmoid forming and propagating within a current sheet ever reported in coronal EUV observations.
Thus, our findings contribute to the large body of observational and numerical efforts regarding plasmoids developed over the past decade.
These include observational plasmoid diagnostics supported by simulations in various eruptive contexts \citep[e.g.,][]{Innes_etal:2015,Rouppe-van-der-Voort_etal:2017,Kumar_etal:2019b,Yan_etal:2022,Cheng_etal:2024}; advanced MHD models capturing plasmoid dynamics \citep[e.g.,][]{Ni_etal:2015,Wyper_etal:2016,Nobrega-Siverio_etal:2017,Hansteen_etal:2019,Ni_etal:2021,Li_etal:2023,Sen_Moreno-Insertis:2025}; as well as ultra-high-resolution \Halpha\ observations revealing the finest-scale signatures of magnetic reconnection \citep[e.g.,][]{Rouppe-van-der-Voort_etal:2023,Kumar_etal:2024}.
The most similar plasmoid scenarios to our observations, combining HRI$_{\mathrm{EUV}}$ and AIA observations, are described by \cite{Mandal_etal:2022}, who reported blobs with sizes between 1-2~Mm within the current sheet of a coronal jet observed at the limb, and by \cite{Cheng_etal:2023}, who identified blob-like features with lifetimes ranging from 5 to 105~s.
The latter referred to them as outflow blobs, which propagate along the fan surface of a CBP and the jet’s outer spine. 
These types of blobs (plasmoids) have also been reported using AIA alone in similar jet contexts \citep[see, e.g.,][]{Zhang_Ji:2014,Zhang_Ni:2019,Kumar_etal:2019b,JoshiR_etal:2020,Mulay_etal:2023}.

The plasmoid analyzed in this paper represents the clearest observational case identified in our dataset.
However, the animation associated with Case 08 (Fig.~\ref{fig:context}) reveals intermittent jetting activity and dynamic behavior of the current sheet, strongly suggesting that multiple plasmoids may be forming and evolving throughout the event, even if only one is clearly resolved.
It is important to emphasize that detecting and tracking such plasmoids requires a delicate balance between high spatial resolution, high temporal cadence, and sufficient exposure time.
In Case 08, the cadence is $\Delta t = 5$~s (see Table~\ref{tab:solo_observations}), which allows us to clearly resolve the small-scale plasmoid in four consecutive frames.
The exposure time is $\mathrm{t}_{\mathrm{exp}} = 1.65$~s, which provides enough signal while still being short enough not to compromise the cadence.
By contrast, when the exposure time drops to $\mathrm{t}_{\mathrm{exp}} = 0.70$~s, as in Case 11, the jet and associated features in the coronal hole become significantly noisier and harder to analyze.
Thus, based on the instrumental configurations of Cases 01 and 08, along with previous similar reports \citep{Mandal_etal:2022,Cheng_etal:2023}, we tentatively suggest that an exposure time in the range $\mathrm{t}_{\mathrm{exp}} \in [1.65, 2.80]$~s and a cadence of $\Delta t \in [3, 5]$~s could provide an optimal balance for studying plasmoid-related signatures in comparable scenarios using HRI$_{\mathrm{EUV}}$.

\subsubsection{Indirect signatures of plasmoids}

Plasmoids cannot always be directly detected in observations due to their scale,  rapid evolution, or because the associated temperature falls outside the bandpass’s response function.
Unlike AIA, HRI$_{\mathrm{EUV}}$ does not observe multiple wavelengths simultaneously at the same resolution, which highlights the need for high-resolution imaging combined with broader temperature diagnostics. 
Still, using a single bandpass, we may find observational signatures in the outflow region that could suggest the presence of plasmoid-mediated reconnection.
This is exemplified in Sect.~\ref{s:indirect_plasmoids}, where we analyze a repeating, boomerang-like pattern with slightly enhanced intensity in the outflow region near the CBP jet of Case 01. 
To support this conjecture, we compare with the numerical simulations, showing that when plasmoids are expelled from the current sheet, they collide with the ambient magnetic field and undergo secondary reconnection, losing their identity as closed magnetic islands.
The plasma originally confined within the plasmoids is redistributed along the newly reconnected field lines.
Because plasmoid-mediated reconnection is inherently bursty, this process creates a sequence of localized density enhancements, separated by lower-density regions.
Given that optically thin emission depends quadratically on the plasma density, this leads to a discontinuous brightness pattern that is consistent with the boomerang-like features observed with HRI$_{\mathrm{EUV}}$.

We cannot completely exclude the possibility that the observed pattern arises from an intermittent reconnection driver rather than from plasmoid ejections. 
However, two key elements lead us to consider plasmoid-mediated reconnection as a plausible scenario in this case.
First, the observed features bear striking similarities to the synthetic HRI$_{\mathrm{EUV}}$ emission derived from the Bifrost simulation, where the presence of plasmoids in density is unambiguous. 
Second, the observed jetting activity of the CBP of Case 01 transitions from a gentle phase to a more explosive one, which could indicate a change in the reconnection regime, from a smooth to a plasmoid-mediated regime, consistent with theoretical expectations for current sheet evolution in high-Lundquist number plasmas.
Moreover, boomerang-like patterns in the outflow region appear to be a generic signature rather than a peculiarity of our simulation. 
Indeed, similar signatures can be seen in synthetic EUV emission from other numerical models with null-point topologies and plasmoid-mediated reconnection, such as figure 6 of \cite{Gannouni_etal:2023} and figure 5 of \cite{Faerder_etal:2024b}, although they are not discussed by the respective authors.
Therefore, these indirect signatures provide valuable insights into the presence of plasmoid-mediated reconnection, complementing other forms of indirect evidence, such as wave-like perturbations along the jet spine \citep[e.g.,][]{Gannouni_etal:2023,HouC_etal:2025}, or quasi-periodic pulsations (QPPs) associated with reconnection dynamics \citep[e.g.,][]{McLaughlin_etal:2018,Kumar_etal:2025}.

\begin{acknowledgements}
We thank the referee for their constructive feedback to improve the presentation and contextualization of the manuscript.
This research has been supported by the European Research Council through the Synergy Grant number 810218 (``The Whole Sun'', ERC-2018-SyG) and by the Research Council of Norway (RCN) through its Centres of Excellence scheme, project number 262622.
D.N.S. and R.J. gratefully acknowledge the Solar Orbiter/EUI Guest Investigator program, and thank Marilena Mierla for her kind support during their two research stays at the Royal Observatory of Belgium.
D.N.S. also acknowledges the support by Aletheia Solaris.
D.L. was supported by a Senior Research Project (G088021N) of the FWO Vlaanderen and the Belgian Federal Science Policy Office through the contract B2/223/P1/CLOSE-UP.
Solar Orbiter is a space mission of international collaboration between ESA and NASA, operated by ESA. The EUI instrument was built by CSL, IAS, MPS, MSSL/UCL, PMOD/WRC, ROB, LCF/IO with funding from the Belgian Federal Science Policy Office (BELSPO/PRODEX PEA 4000112292 and 4000134088); the Centre National d’Etudes Spatiales (CNES); the UK Space Agency (UKSA); the Bundesministerium für Wirtschaft und Energie (BMWi) through the Deutsches Zentrum für Luft- und Raumfahrt (DLR); and the Swiss Space Office (SSO).
The authors also acknowledge the computer resources at the MareNostrum
supercomputing installation and the technical support provided by the
Barcelona Supercomputing Center (BSC, RES-AECT-2021-1-0023,
RES-AECT-2022-2-0002), as well as the resources provided by 
Sigma2 - the National Infrastructure for High Performance Computing and 
Data Storage in Norway.
\end{acknowledgements}

\bibliographystyle{aa}
\bibliography{references}

\appendix
\onecolumn
\section{Context from SDO: AIA and HMI}\label{app:sdo}
\begin{figure*}
   \centering
   \includegraphics[width=0.80\textwidth]{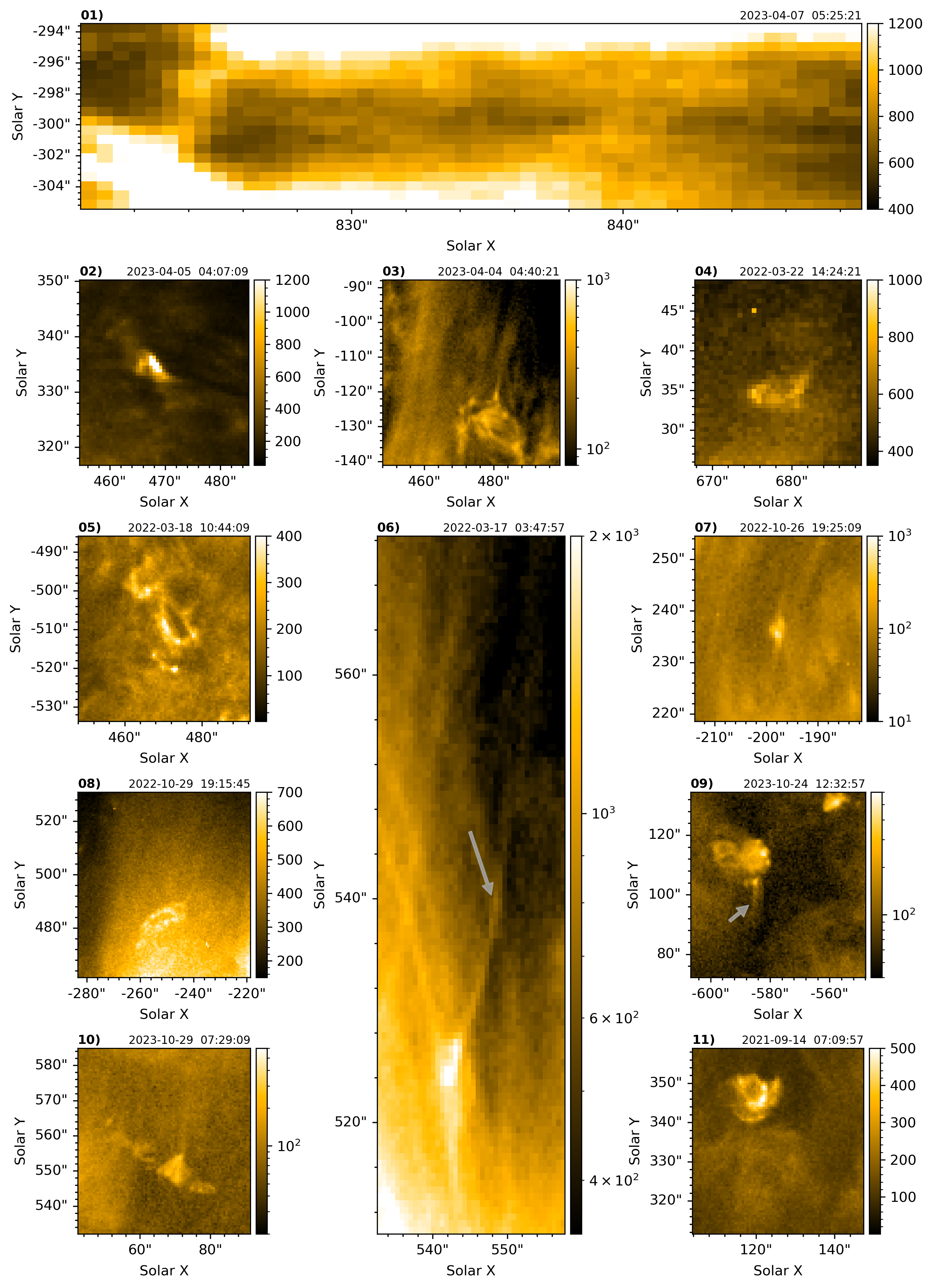}
   \caption{SDO/AIA 171~\AA\ observations corresponding to the eleven thin coronal jets associated with CBPs shown in Fig.~\ref{fig:context}.
   Cases 06 and 09 are those in which the spine of the jet, indicated by the gray arrows, can be easily distinguished in AIA.
   In the other cases, the jet base may be inferred, or the jet is completely absent.
   The intensity of the images is given in DN and times are in UT.
   The images display the same regions as observed by HRI$_{\mathrm{EUV}}$ but are not spatially co-aligned, as the goal is to provide contextual information. 
   The AIA times have been selected to match the corresponding HRI$_{\mathrm{EUV}}$ observations, accounting for the light travel time delay due to Solar Orbiter’s position near the Sun.
   Individual movies for each dataset are available \href{https://zenodo.org/records/16903189}{online}}.
   \label{fig:appendix_aia}
\end{figure*}

\begin{figure*}
   \centering
   \includegraphics[width=0.80\textwidth]{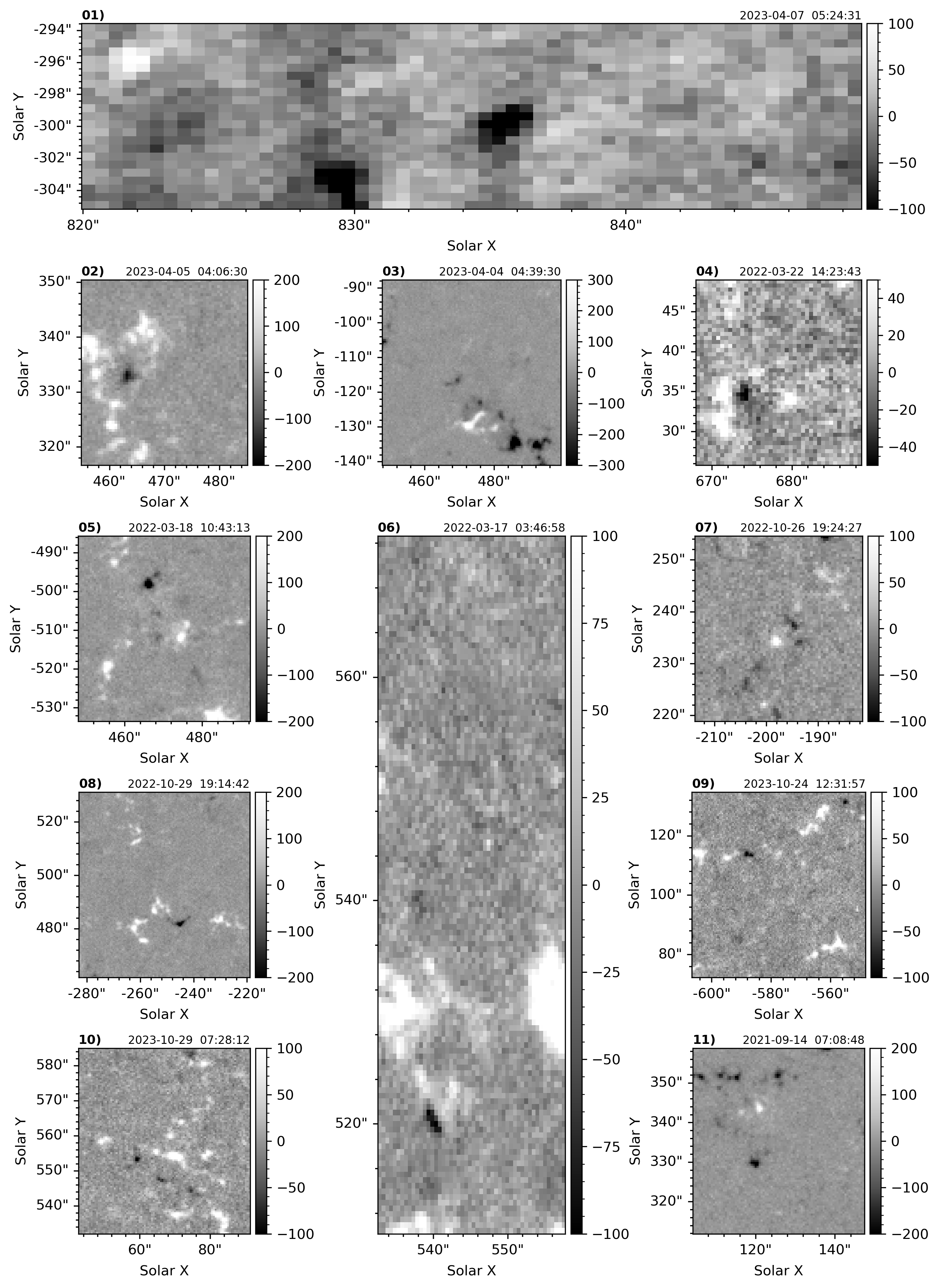}
   \caption{Line-of-sight photospheric magnetograms from SDO/HMI corresponding to the eleven thin coronal jets associated with CBPs discussed in this paper.
   The field strength of the images is given in G and times are in UT.
   The HMI maps were acquired at the time closest to the AIA images shown in Fig. \ref{fig:appendix_aia}, and have been co-aligned accordingly.
   }
   \label{fig:appendix_hmi}
\end{figure*}

Figure~\ref{fig:appendix_aia} presents SDO/AIA 171~\AA\ observations of the same regions shown in Fig.~\ref{fig:context}.
The AIA times were selected to match the corresponding HRI$_{\mathrm{EUV}}$ observations, accounting for the light travel time delay due to Solar Orbiter’s proximity to the Sun.
Since our goal here is purely illustrative, the AIA maps are not spatially co-aligned with the HRI$_{\mathrm{EUV}}$ data.
Nonetheless, the main large-scale structures can still be identified in Fig.~\ref{fig:appendix_aia}, confirming that we are covering the same regions.
In some cases (e.g., Cases 06 and 09), the narrow jet can be discernible in AIA and extends over several megameters (see arrows in the plots). 
In others, its presence can be inferred based on the HRI$_{\mathrm{EUV}}$, especially close to the jet base (e.g., Cases 02, 07, 08, and 10).
In Case 05, for instance, only the base of the narrowest jet is discernible in AIA, even though other jetting episodes from the same CBP are detectable (see associated animation).
In the most extreme examples (e.g., Cases 01 and 11), the jet is not visible in AIA.
To provide additional context, in Fig.~\ref{fig:appendix_hmi} we include line-of-sight photospheric magnetograms from SDO/HMI, taken at the closest time to the AIA observations and co-aligned accordingly.
In all cases except Case 01, we can easily discern one or more parasitic polarities embedded within an opposite-sign background field, 
a typical magnetic configuration leading to CBPs \citep[e.g.,][]{Zhang_etal:2012, Mou_etal:2016, Galsgaard_etal:2017, Madjarska_etal:2021}.

\FloatBarrier
\newpage

\section{Length and width of the jets}\label{app:width}
\begin{figure*}
   \centering
   \includegraphics[width=0.90\textwidth]{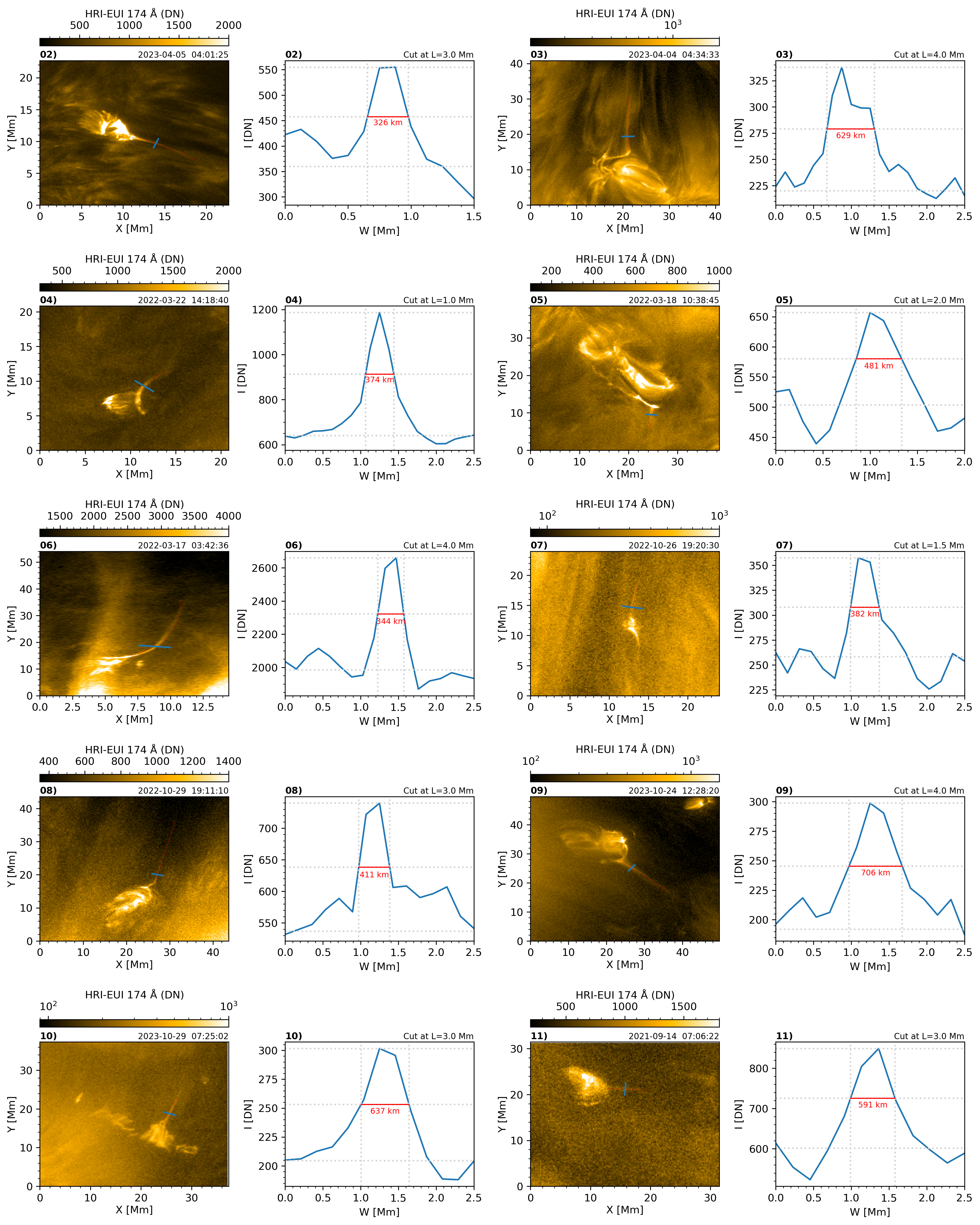}
   \caption{Widths of narrow coronal jets emanating from CBPs for Cases 02 to 11.
   The red dashed line in the HRI$_{\mathrm{EUV}}$ images indicates the Bézier curve that fits the spine of the jet, while the perpendicular blue line marks the cut used to determine the FWHM shown in the adjacent panel.
   In the 1D plots, the dotted gray lines serve as visual aids to illustrate how the FWHM is determined.
   }
   \label{fig:appendix_width}
\end{figure*}

To determine the length $L$ of the coronal jet, we employ a quadratic Bézier curve, thus getting a smooth and continuous curve along the jet.  
The curve is parameterized by $s \in [0,1]$ and is given by:
\begin{equation}
  \mathbf{L}(s) = (1 - s)^2 \mathbf{P}_0 + 2(1 - s)s \mathbf{P}_1 + s^2 \mathbf{P}_2,  
\end{equation}
where $\mathbf{P}_0$, $\mathbf{P}_1$, and $\mathbf{P}_2$ are the position vectors of the base of the jet, a central control point, and the top of the jet, respectively. 
The width is then obtained by computing the full width at half maximum (FWHM) of a perpendicular cut at a given distance along the jet. 
The intensity threshold for the FWHM is set as the average of the values at the endpoints of the perpendicular cut.
The results for Cases 02 to 11 are shown in Fig.~\ref{fig:appendix_width}.
The figure also demonstrates that the brightness of these coronal jets from CBPs ranges from 30\% (Case 05) to 85\% (Case 04) above that of the surrounding emission.

\end{document}